\documentclass[aps,pre,superscriptaddress,twocolumn]{revtex4-1}

\usepackage{graphicx}

\usepackage[utf8]{inputenc}
\usepackage[english]{babel}
\usepackage{amsmath,amssymb,enumerate}

\usepackage[breaklinks]{hyperref}

\usepackage{bm}

\newcommand{\st}{\text{s}}

\newcommand{\rev}{\text{rev}}
\newcommand{\irr}{\text{irr}}
\newcommand{\bath}{\text{b}}

\begin{document}


\title{Derivation of a Langevin equation in a system with multiple scales: the case of negative temperatures} 


\author{Marco Baldovin}
\affiliation{Dipartimento di Fisica,  Universit\`a di Roma Sapienza, P.le Aldo Moro 2, 00185 Roma, Italy}
\author{Angelo Vulpiani}
\affiliation{Dipartimento di Fisica,  Universit\`a di Roma Sapienza,
  P.le Aldo Moro 2, 00185 Roma, Italy}
\author{Andrea Puglisi}
\affiliation{Istituto dei Sistemi Complessi - CNR and Dipartimento di Fisica, Universit\`a di Roma Sapienza, P.le Aldo Moro 2, 00185, Rome, Italy}
\author{Antonio Prados}
\affiliation{F\'isica Te\'orica, Universidad de Sevilla, Apartado de  Correos 1065, E-41080 Sevilla, Spain}
\email[]{prados@us.es}


\date{\today}

\begin{abstract}
  We consider the problem of building a continuous stochastic model,
  i.e. a Langevin or Fokker-Planck equation, through a well-controlled
  coarse-graining procedure. Such a method usually involves the
  elimination of the fast degrees of freedom of the ``bath'' to which
  the particle is coupled. Specifically, we look into the general case
  where the bath may be at negative temperatures, as found--for
  instance--in models and experiments with bounded effective kinetic
  energy. Here, we generalise previous studies by considering the case
  in which the coarse-graining leads to (i) a renormalisation of the
  potential felt by the particle, and (ii) spatially dependent
  viscosity and diffusivity. In addition, a particular relevant
  example is provided, where the bath is a spin system and a sort of
  phase transition takes place when going from positive to negative
  temperatures. A Chapman-Enskog-like expansion allows us to
  rigorously derive the Fokker-Planck equation from the microscopic
  dynamics. Our theoretical predictions show an excellent agreement
  with numerical simulations.
\end{abstract}


\maketitle

\textit{Introduction.-} Systems with negative temperature typically
appear in experiments or models where the effective kinetic and
potential energies are limited and therefore the microcanonical
entropy can be non-monotonic in the
energy~\cite{landau_statistical_2013,dunkel_consistent_2014,vilar_communication:_2014,frenkel_gibbs_2015,puglisi_temperature_2017}. Examples
are found in many physical contexts, including nuclear
spins~\cite{purcell_nuclear_1951,hakonen_negative_1994}, fluid
dynamics~\cite{onsager_statistical_49} and trapped ultra-cold
atoms~\cite{rapp_equilibration_2010,braun_negative_2013}. In these
systems, the presence of negative temperatures is seen without
ambiguities when observing certain degrees of freedom: for instance
the single particle momentum distribution may take the typical form of
an ``inverted'' Maxwell-Boltzmann distribution, of course with cut-off
values at the boundaries~\cite{cerino_consistent_2015}.

It is worth recalling that negative values of temperature-like
variables also arise in other physical frameworks, for example, within
Edwards's statistical mechanics description of dense granular media~
\cite{edwards_theory_1989,mehta_statistical_1989,baule_edwards_2018}. Therein,
the role analogous to that of the temperature is played by the
compactivity $X$, which is defined by
$X^{-1}=\partial S(V)/\partial V$, where $S(V)$ is the total number of
stable configurations for a given volume $V$. Since $S(V)$ is not a
monotonic function of $V$, negative compactivities arise and
correspond to packings that are looser than those characterised by
positive values of $X$
~\cite{brey_thermodynamic_2000,brey_closed_2003,ciamarra_random_2008,briscoe_jamming_2010}.

Once the thermodynamics and the statistical mechanics of a class of
systems has been understood, it is a natural question to wonder about
their (statistical) dynamical description. A classical problem is that
of deducing stochastic equations for the dynamics of slow degrees of
freedom, for example a Langevin equation (LE) for the evolution of
the position and/or momentum of a tagged massive
particle~\cite{zwanzig_nonlinear_1973}. In the following, by LE we
mean a stochastic differential equation, which corresponds to a
continuous Markov process~\cite{van_kampen_stochastic_1992}. It is
important to recall that analytical derivations thereof, through some
kind of coarse-graining procedures, from the equations of the
``microscopic'' dynamics--e.g. Hamilton equations, Liouville equation,
Boltzmann equation, etc.--are possible only in few special cases. A
relevant alternative is to assume some form of LE with few parameters,
based upon some previous theoretical knowledge of the investigated
problem, and then estimate those parameters from numerical or
experimental data through a proper inferring procedure. A discussion
of such an approach and its many practical subtleties is given
in Ref.~\cite{baldovin_langevin_2018}.

In the case of systems with negative temperature, a LE for a massive
particle has already been considered by some of us
in~\cite{baldovin_langevin_2018}. Therein, it was assumed that the
parameters appearing in the LE--viscosity and noise amplitude--were
constant. In general, however, it may happen that there is a coupling
of the transport coefficients of the LE with the
particle position, depending upon the particular form of the global
Hamiltonian. Moreover, in such a previous investigation, a procedure
to infer the viscosity--or noise amplitude--from the Hamiltonian of
the total system was not provided: on the contrary, it was shown the
fair success of an inference recipe of LE parameters from numerical
data.

In the present paper, we consider a more general case that includes,
in addition to possible negative temperatures, (i) a renormalisation
of the potential felt by the heavy particle, and (ii) inhomogeneous LE
parameters. The usual Einstein-like relation between viscosity and
noise amplitude is confirmed by simply assuming equilibrium, with the
particular form of kinetic energy not playing any crucial
role. Afterwards, as an example, we investigate a Hamiltonian system
comprising a slow continuous degree of freedom coupled to a bath of
spins. A Chapman-Enskog-like coarse-graining procedure allows us to
derive the LE for the slow degrees of freedom, which leads to
both a renormalised potential and non-uniform viscosity and noise
amplitudes, obeying the Einstein relation mentioned
before. Interestingly, a phase transition--in a sense to be specified
below--stems from the renormalisation of the potential, when the
temperature crosses from positive to negetive values. Numerical
simulations of the total Hamiltonian systems and the LE
confirm our theoretical picture.

\textit{Renormalised potential and generalised Einstein-relation
  between viscosity and diffusivity.-} Let us consider a system
comprising a ``heavy'' particle with canonical variables
$\Gamma\equiv(x,p)$ and a bath characterised by some variables that we
denote by $\bm{z}$.
The Hamiltonian of this system is assumed to have the form
\begin{equation} \label{ham_gen}
H(\Gamma,\bm{z})=K(p)+U(x)+H_N(\bm{z})+ V_I(x,\bm{z}),
\end{equation}
where $K$ and $U$ are the ``kinetic energy'' of the slow particle and its
external confining potential, respectively, $H_N$ is the Hamiltonian
of the bath, and finally $V_I$ is the potential for the interaction
between the heavy particle and the bath. The bath variables $\bm{z}$
can be, for example, positions and momenta of $N$ ``light''' particles
or Ising variables of $N$ ``fast'' spins.

At equilibrium at temperature $T$, the probability distribution
function (PDF) for the whole system is given by the canonical
distribution,
$\mathcal{P}_{\st}(\Gamma,\bm{z})=Z^{-1}\exp[-\beta
H(\Gamma,\bm{z})]$, where $Z$ is the partition function and $\beta$ is
the inverse of the temperature--we are taking Boltzmann's constant
$k_{B}=1$. The marginal PDF for the particle variables is then given
by
\begin{subequations}
\begin{equation}
  \label{eq:marginal-prob-eq}
  f_{\st}(\Gamma)=Z^{-1}e^{-\beta[K(p)+U_{R}(x)]},
  \quad U_{R}(x)=U(x)+\mathcal{F}_{\bath}(x),
\end{equation}
\begin{equation}
  \label{eq:renorm-potential}
  e^{-\beta \mathcal{F}_{\bath}(x)}\equiv\int d\bm{z}\, e^{-\beta[H_N(\bm{z})+ V_I(x,\bm{z})]}.
\end{equation}
\end{subequations}
Note that, in general, the integration over the bath variables
renormalises the potential felt by the particle. The additional term
$\mathcal{F}_{\bath}(x)$ is the free energy of the bath for given
values of the particle variables.

Now we turn our attention to the dynamics. The evolution equations for
$(x,p)$ read
\begin{subequations}
  \begin{align}
    \dot{x}&=\partial_p H=K^{\prime}(p)\\
    \dot{p}&=-\partial_x H=-U^{\prime}(x)-\partial_x V_I(x,\bm{z}), \label{dotP}
  \end{align}
\end{subequations}
where the prime indicates the relevant derivative for functions that
only depend on one variable. At this point, we introduce the
hypothesis of time-scale separation: the heavy particle variables
$(x,p)$ are assumed to evolve much slower than the bath variables
$\bm{z}$. In this regime, the term $\partial_x V_I$ is expected to be
described by a ``viscous term''--only function of $(x,p)$--plus a
``noisy term''. In other words, we seek to generalise the
Klein-Kramers equation to systems with a generic form of $K(p)$, which
may allow for the existence of negative temperatures.

Following the above discussion, our candidate equation has the generic
form
\begin{equation} \label{KK}
  \dot{x} = K^{\prime}(p), \qquad
  \dot{p} = -U^{\prime}(x) + B(\Gamma,t),
\end{equation}
in which $B(\Gamma,t)$ is the effective force--which contains also a noisy term--on the particle stemming
from the interaction with the bath.
Going from Eq.~\eqref{dotP} to~\eqref{KK} 
implies conditional averages over the fast degrees of freedom, keeping
fixed the slow variables. Therefore, the statistical
properties of the coarse-grained force $B(\Gamma,t)$ depend in general on
both $x$ and $p$.

The original Langevin-Klein-Kramers equation equation predicts a
linear--or additive--form for $B$, namely
$B(\Gamma,t)=-\gamma p + \sqrt{2D}\xi(t)$.
Therein, $\xi(t)$ is a Gaussian white noise, with
$\langle \xi(t) \rangle=0$ and
$\langle \xi(t)\xi(t') \rangle = 2D\delta(t-t')$, and the two main
parameters are the constant viscosity $\gamma>0$ and diffusivity
$D>0$.  When $K(p)$ is not quadratic in $p$, the simplest modification
is replacing the viscous term $-\gamma p$ with
$- \gamma K^{\prime}(p)$. This was done
in~\cite{baldovin_langevin_2018}, where the usual Einstein relation
$\gamma=\beta D$ was shown to hold also for $\beta<0$.

The coarse-graining over the bath variables may lead to a more general
situation, which we analyse here. First, an additional effective
external potential term may appear in $B(\Gamma,t)$, which we identify
with $-\mathcal{F}'_{\bath}(x)$ to be consistent with the equilibrium
situation: if the bath variables were infinitely fast, the bath would
remain exactly at equilibrium at all times and the particle would
follow a deterministic motion under the force
$-U'_{R}(x)=-U'(x)-\mathcal{F}'_{\bath}(x)$~\footnote{Note that
  $\mathcal{F}'_{b}(x)$ equals the average value of
  $\partial_{x}V_{I}(x,\bm{z})$ when the fast variables $\bm{z}$ are
  at equilibrium at a fixed value of $x$, as predicted by
  \eqref{eq:renorm-potential}.}. Second, the viscosity and the
diffusivity may be spatially dependent, i.e $\gamma=\gamma(x)$ and
$D=D(x)$. Incorporating these two ingredients into our description, we
end up with the following ansatz for the coarse-grained force
\begin{equation} \label{KK2}
  B(\Gamma,t) = -\mathcal{F}_{\bath}^{\prime}(x)-\gamma(x)K^{\prime}(p) +\sqrt{2D (x)}\xi(t).
\end{equation}

The Fokker-Planck equation for the PDF $f(\Gamma,t)$ for the heavy
particle is then
\begin{align}
  \partial_{t}f=&-K^{\prime}(p)\partial_{x}f+U_{R}^{\prime}(x)
                  \partial_{p} f\nonumber \\ &+\partial_{p}\left[\gamma(x)K^{\prime}(p)f+D(x)\partial_{p}f\right], \label{eq:F-P-generic}
\end{align}
Following \cite{risken_solutions_1972}, we can write the Fokker-Planck
equation as a conservation law, $\partial_{t}f=-\nabla\cdot\bm{J}$,
where $\bm{J}\equiv\{J_{x},J_{p}\}$ is the probability density
current and
$\nabla\cdot\bm{J}\equiv\partial_{x}J_{x}+\partial_{p}J_{p}$. Moreover,
$\bm{J}$ can be split into its reversible and irreversible parts
$\bm{J}_{\rev}$ and $\bm{J}_{\irr}$, specifically
$\bm{J}_{\rev}(\Gamma,t)= \{
K^{\prime}(p)f(\Gamma,t),-U_{R}^{\prime}(x)f(\Gamma,t)\}$ and
$\bm{J}_{\irr}(\Gamma,t)= \{0,-\gamma(x)
K^{\prime}(p)f(\Gamma,t)-D(x)\partial_pf(\Gamma,t)\}$.


The steady solution of the Fokker-Planck equation must be the
equilibrium distribution $f_{\st}(\Gamma)$ in
Eq.~\eqref{eq:marginal-prob-eq}. On the one hand, substitution of the
steady distribution into the Fokker-Planck equation always leads to
$\nabla \cdot \bm{J}_{\rev,\st}(\Gamma)\equiv 0$, with no particular
requirements for the reversible part of the current.  On the other
hand, the condition $\nabla\cdot\bm{J}_{\irr,\st}=0$ can be fulfilled
only if $\bm{J}_{\irr,\st}\equiv 0$, i.e., if detailed balance (DB)
holds~\cite{risken_solutions_1972}.  The DB condition leads to
\begin{equation}\label{eq:Einstein-general}
  \gamma(x)=\beta D(x),
\end{equation}
which is a
generalised Einstein relation for inhomogeneous viscosity and
diffusivity. 

\textit{An example with analytical derivation of the LE} - As an
example of the general case discussed before, we consider the
following Hamiltonian for a slow particle  coupled to a spin
bath,
\begin{subequations}\label{eq:1}
  \begin{align}
    H(\Gamma, \bm{\sigma})&=K(p)+V(x,\bm{\sigma}), \\
    V(x,\bm{\sigma})&=U(x)-\mu \lambda(x)
    \sum_{j=1}^N\sigma_{j}.
  \end{align}
\end{subequations}
Above, $\bm{\sigma}\equiv(\sigma_1, \sigma_2, ..., \sigma_N)$ are spin
variables, $\sigma_{j}=\pm 1$, $\mu$ is a constant, and $\lambda(x)$
is a certain function of $x$. Then, the spins $\bm{\sigma}$ are
the bath variables $\bm{z}$ in Eq.~\eqref{ham_gen}, and the bath
contribution to the Hamiltonian $H_{N}({\bm{z}})+V_{I}({x,\bm{z}})$
reduces to the term $-\mu\lambda(x)\sum_{j}\sigma_{j}$, i.e. the spins
feel an inhomogeneous external field $\mu\lambda(x)$.

To start with, we discuss the equilibrium situation. Therein, the
system as a whole is described by the canonical distribution
$ \mathcal{P}_{\st}(\Gamma,\bm{\sigma})=Z^{-1}\exp\left[-\beta
  H(\Gamma,\bm{\sigma})\right]$. In this simple case, the specific
form of the free energy of the bath $\mathcal{F}_{\bath}(x)$ for given
values of the particle variables is
\begin{equation}\label{eq:Z-spins}
e^{-\beta\mathcal{F}_{\bath}(x)}=\left\{ 2\cosh\left[\beta \mu \lambda(x)\right]\right\}^{N}.
\end{equation} 
Moreover, we can also write the conditional probability of finding the
spins in a configuration $\bm{\sigma}$ for given values of the
particle variables as
\begin{equation}
\label{eq:10}
 \mathcal{P}^{\st}(\bm{\sigma}|x) = e^{\beta \left[\mu \lambda(x) \sum_{j} \sigma_{j}+\mathcal{F}_{\bath}(x)\right]}.
\end{equation}
Our notation makes it explicit that this conditional probability
depends only on $x$. Also, we have that
\begin{equation}
  \label{eq:16}
  \mathcal{F}'_{\bath}(x)=-N\mu\lambda'(x)\langle\sigma\rangle_{\st}(x),
  \quad \langle \sigma \rangle_{\st}(x) = \tanh\left[ \beta \mu \lambda(x) \right]\,,
\end{equation}
where
$\langle \sigma \rangle_{\st}(x)\equiv\sum_{\bm{\sigma}}
\sigma_{j}\mathcal{P}^{\st}(\bm{\sigma}|x)$, for any $j$.

Now, let us consider the dynamics. On the one hand,
accordingly with our previous general discussion, the evolution
equations for $(x,p)$ are
\begin{equation}
\label{eq:2} \dot{x}=\frac{\partial H}{\partial p}= K'(p), \quad 
\dot{p}=-\frac{\partial H}{\partial x}=-\partial_{x}V(x,\bm{\sigma}),
\end{equation}
where
$\partial_{x}V(x,\bm{\sigma})=
U'(x)-\mu\lambda'(x)\sum_{j}\sigma_{j}$.
On the other hand, and for the sake os simplicity, we assume Glauber's
stochastic dynamics for the spins. We denote by $\mathcal{R}_{j}$ the
operator that flips the $j$-th spin, leaving the remainder
unchanged.  The transition rate for the flipping of the $j$-th spin,
i.e. from configuration $\bm{\sigma}$ to $\mathcal{R}_{j}\sigma$,
is
\begin{equation}
\label{eq:3}
 W_{j}(\bm{\sigma}|x)=\frac{\alpha}{2} \left\{ 1- \sigma_{j}\tanh[\beta \mu \lambda(x)] \right\},
\end{equation}
in which $\alpha$ is a characteristic
rate~\cite{glauber_time-dependent_1963}. We can write a
Liouville-master equation for the time evolution of the joint
PDF $\mathcal{P}(\Gamma,\bm{\sigma},t)$,
\begin{align}
\label{eq:4}
  \mathbb{W}(\bm{\sigma}|x)\mathcal{P}(\Gamma,\bm{\sigma},t)
     =\varepsilon\left[ \partial_t + \mathcal{L}(\Gamma,\bm{\sigma})\right]\mathcal{P}(\Gamma,\bm{\sigma},t)\,.
\end{align}
We have introduced the linear operators
\begin{subequations}
\begin{align}
  \mathbb{W}(\bm{\sigma}|x)&\equiv\sum_{j=1}^N
  (\mathcal{R}_j-1)W_j(\bm{\sigma}|x), \\
  \mathcal{L}(\Gamma,\bm{\sigma})&\equiv K'(p)\partial_x
  -\partial_{x}V(x,\bm{\sigma}) \partial_p.
\end{align}
\end{subequations}
Note the auxiliary $\varepsilon$ in front of the right-hand side (rhs)
of Eq.~\eqref{eq:4}, actually $\varepsilon=1$. Clearly, the canonical
distribution is a time-independent solution of Eq.~\eqref{eq:4}~\footnote{See Appendix A of
  \cite{bonilla_nonequilibrium_2010} for a proof of an $H$-theorem for
  this kind of system, specifically for quadratic $K(p)$, $U(x)$ and
  linear $\lambda(x)$, although these particular shapes are not
  required in the proof.}.

Our idea is to derive an equation for the marginal PDF for the
particle variables
$f(\Gamma,t)=\sum_{\bm{\sigma}}\mathcal{P}(\Gamma,\bm{\sigma},t)$
when the spins are much faster than the ``heavy''
particle. Specifically, this means that $\omega_0/\alpha\ll 1$, with
$\omega_{0}^{-1}$ being the characteristic time over which the
``heavy'' particle evolves. Instead of making this idea explicit by
introducing dimensionless variables, we have employed an equivalent
approach--usual in kinetic theory--by introducing the auxiliary
$\varepsilon$ in front of the rhs of Eq.~\eqref{eq:4}~\footnote{In
  dimensionless variables, we would have $\delta= \omega_0/\alpha$ in
  front of the rhs; thus we are making an expansion in powers of
  $\delta$.}.

\textit{Chapman-Enskog expansion.-} We proceed with an expansion in powers of $\varepsilon$,
\begin{equation}
\label{eq:9}
 \mathcal{P}(\Gamma,\bm{\sigma},t) = \mathcal{P}^{\st}(\bm{\sigma}|x)f(\Gamma,t)+\sum_{l=1}^{\infty} \varepsilon^l \,\mathcal{P}^{(l)}(\Gamma,\bm{\sigma},t).
\end{equation}
We ensure $f(\Gamma,t)$ to be the exact marginal distribution of the
particle by assuming
$\sum_{\bm{\sigma}} \mathcal{P}^{(l)}(\Gamma,\bm{\sigma},t) =0$,
$\forall l \ge 1$.
It is the dynamical equation of $f(x,p,t)$--and not $f$ itself--that
is expanded in powers of $\varepsilon$ in the Chapman-Enskog
method~\cite{resibois_classical_1977,bonilla_chapman-enskog_2000,bonilla_nonlinear_2009,neu_singular_2015,bonilla_unpublished_2019}
\begin{equation}
\label{eq:12}
 \partial_t f(\Gamma,t) =\sum_{l=0}^{\infty} \varepsilon^l F^{(l)}(\Gamma,t)\,.
\end{equation}
Truncating the above series at the lowest order ($l=0$), one has the
``deterministic'' (zero noise) approximation. The effect of the noise
can be introduced in the simplest way by retaining the first two terms
($l=0,1$). This is what we do in the
following~\footnote{$F^{(l)}(\Gamma,t)$ is a notation we use to stress
  that these functions do not depend on $\bm{\sigma}$ but on
  $(\Gamma, t)$ both directly and indirectly through $f(\Gamma,t)$ and
  its derivatives.}.

Now, we list the equations obtained by inserting Eqs.~\eqref{eq:10},
\eqref{eq:9}, and \eqref{eq:12} into Eq.~\eqref{eq:4}. Up to order
$\varepsilon^{2}$,
\begin{subequations}
 \begin{align}
 \label{eq:13a}
 &   \mathbb{W}(\bm{\sigma}|x) \mathcal{P}^{\st}(\bm{\sigma}|x)
   f(\Gamma,t) = 0, \\  
 &   \mathbb{W}(\bm{\sigma}|x) \mathcal{P}^{(1)}(\Gamma,\bm{\sigma},t)
   = \mathcal{P}^{\st}(\bm{\sigma}|x)F^{(0)}(\Gamma,t) \nonumber\\ 
 \label{eq:13b}    
&  \qquad\qquad\qquad\qquad\qquad + \mathcal{L}(\Gamma,\bm{\sigma}) \mathcal{P}^{\st}(\bm{\sigma} | x) f(\Gamma, t),\\
 \label{eq:13c}    
 &   \mathbb{W}(\bm{\sigma}|x)
                      \mathcal{P}^{(2)}(\Gamma,\bm{\sigma},t)  = 
                      \mathcal{P}^{\st}(\bm{\sigma}|x)F^{(1)}(\Gamma,t)
                      \nonumber \\
           &   \quad\qquad\qquad\qquad\qquad\quad+ \left[ \partial_{t}+\mathcal{L}(\Gamma,\bm{\sigma})\right]
                   \mathcal{P}^{(1)}(\Gamma,\bm{\sigma},t)\,.
 \end{align}
\end{subequations}
Equation~\eqref{eq:13a} (order of unity, $\varepsilon^{0}$) is an
identity, because in Eq.~\eqref{eq:9} we have anticipated the zero-th
order contribution to the expansion of $\mathcal{P}$ in powers of
$\varepsilon$.

First, we resort to Eq.~\eqref{eq:13b} (order of $\varepsilon$) to obtain
$F^{(0)}$ and $\mathcal{P}^{(1)}$. We bring to bear that the rhs of
Eq.~\eqref{eq:13b} must be orthogonal to
$\mathcal{P}^{\st}(\bm{\sigma}|x)$, i.e. its sum over all the spin
configurations vanishes, which entails that
$F^{(0)}=-K'(p)\partial_x f+ \left[
  U'(x)+\mathcal{F}'_{\bath}(x)\right]\partial_p f$.
Following our general discussion, there appears an extra force
$-\mathcal{F}'_{\bath}(x)$, given in this specific system by
Eq.~\eqref{eq:16}. In order to have a consistent limit as $N \to \infty$,
the coupling constant $\mu$ between the particle and the spins must
scale as $N^{-1/2}$~\footnote{This is a typical scaling for the
  coupling constant between the heavy particle and the ``bath'', see
  for instance \cite{zwanzig_nonlinear_1973} for the classical case of
  a Brownian particle coupled to a bath of harmonic
  oscillators.}. With this scaling, we have that
\begin{subequations}
 \label{eq:21}
 \begin{alignat}{2}
 \label{eq:21a} F^{(0)}(\Gamma,t)&=-K'(p)\partial_x f(\Gamma,t) + U'_R(x) \partial_p f(\Gamma,t)\\
 \label{eq:21b} U_R(x)&=U(x)-\frac{\beta}{2}\tilde{\mu}^2
 \lambda^2(x)\,, \quad \mu = \tilde{\mu} N^{-1/2},
 \end{alignat}
\end{subequations}
It is worth emphasising the emergence of the ``renormalised''
potential $U_R(x)$, once more accordingly with the general framework
developed before. 

Next,  we substitute the
obtained expressions for $F^{(0)}$ into Eq.~\eqref{eq:13b}, and
take into account that $\partial_x \mathcal{P}^{\st}(\bm{\sigma}|x) = \beta \mu \lambda '(x) \sum_j \left[\sigma_j -\langle \sigma \rangle_{\st}(x)\right]\mathcal{P}^{\st}(\bm{\sigma}|x)$
to write the following equation for $\mathcal{P}^{(1)}$,
\begin{align}
 \label{eq:23}
\!\!\!\! \mathbb{W}(\bm{\sigma}|x)
                        \mathcal{P}^{(1)}(\Gamma,\bm{\sigma},t)&= \mu \lambda '(x) \left[ \beta K'(p) f(\Gamma,t) + \partial_p f(\Gamma,t)\right] \nonumber
  \\
  &  \quad\times \sum_{j=1}^{N}\left[\sigma_{j} - \langle \sigma\rangle_{\st}(x)\right]\mathcal{P}^{\st}(\bm{\sigma}|x)\,.
\end{align} 
Interestingly, this equation can be explicitly solved for
$\mathcal{P}^{(1)}$, because it is easy to show that
$\sum_{j}\left[\sigma_{j} - \langle
  \sigma\rangle_{\st}(x)\right]\mathcal{P}^{\st}(\bm{\sigma}|x)$ is an
eigenvector of the operator $\mathbb{W}(\bm{\sigma}|x)$ corresponding
to the eigenvalue $-\alpha$. Therefore,
\begin{align}
 \mathcal{P}^{(1)}(\Gamma,\bm{\sigma},t)&=-\alpha^{-1}\mu\lambda'(x)\left[
    \beta K'(p)f(\Gamma,t)+\partial_{p}f(\Gamma,t)\right] \nonumber \\
  & \quad\times\sum_{j=1}^{N}\left[\sigma_{j} - \langle
    \sigma\rangle_{\st}(x)\right]\mathcal{P}^{\st}(\bm{\sigma}|x).
\end{align}

Now, we make use of Eq.~\eqref{eq:13c} to calculate
$F^{(1)}$~\footnote{Note that $\mathcal{P}^{(2)}$ would only be
  necessary if we were interested in higher order terms in the
  equation for $\partial_t f$, such as $F^{(2)}$.}: its rhs must also
be orthogonal to $\mathcal{P}^{\st}(\bm{\sigma}|x)$, i.e. the sum over
all the spin configurations must vanish. Therefore,
$F^{(1)}=- \mu \lambda'(x) \sum_{\bm{\sigma}}\sum_{j} \sigma_{j}
\partial_p \mathcal{P}^{(1)}$, from which (i) taking into account the
explicit expression for $\mathcal{P}^{(1)}$ and (ii) considering
the limit as $N\to \infty$, $F^{(1)}$ is reduced to
\begin{equation}
 \label{eq:30}
 F^{(1)}(\Gamma,t)=\alpha^{-1}\left[ \tilde{\mu} \lambda'(x) \right]^2 \partial_p \left[ \beta K'(p) f(\Gamma,t) + \partial_p f(\Gamma,t)\right]\,.
\end{equation}

\textit{Fokker-Planck equation for $f(\Gamma,t)$.-} Up to order
$\varepsilon$, the evolution of the marginal distribution
$f(\Gamma,t)$, is given by
$\partial_t f =F^{(0)}+ \varepsilon F^{(1)}$. We write the result in
the limit as $N \to \infty$, with the scaling in Eq.~\eqref{eq:21b}
and, moreover, we make $\varepsilon=1$ as we discussed before carrying
out the Chapman-Enskog expansion. Making use of
Eqs.~\eqref{eq:21}~and~\eqref{eq:30}, we arrive at
\begin{align}
 \label{eq:32}
 \partial_t f &= -K'(p)\partial_x f + U'_R(x) \partial_p f \nonumber
  \\
  & \quad + \alpha^{-1}\left[\tilde{\mu} \lambda'(x)\right]^2 \partial_p \left[ \beta K'(p) f + \partial_p f \right]\,
\end{align}
which is in complete agreement with the general picture we
have developed before. In particular, comparison with
Eq.~\eqref{eq:F-P-generic} leads to identifying the viscosity and the
diffusivity in terms of the microscopic parameters of the model,
\begin{equation}
  \label{eq:39}
  D(x)=\alpha^{-1}\left[\tilde{\mu}\lambda'(x)\right]^{2}, \quad
  \gamma(x)=\beta D(x).
\end{equation}
Of course, the stationary solution of Eq.~\eqref{eq:32} is the exact
marginal equilibrium distribution
$f_{\st}(\Gamma)\propto e^{-\beta \left[ K(p) +U_R(x) \right]}$, in
accordance with Eq.~\eqref{eq:marginal-prob-eq}.

The Fokker-Planck equation \eqref{eq:32} can be rewritten as a
LE,
\begin{equation}
 \label{eq:37}
  \dot{x}=K'(p), \quad 
 \dot{p}=-U'_R(x)-\alpha^{-1}\left[ \tilde{\mu} \lambda'(x)\right]^2 \beta K'(p) + \xi(t),
\end{equation}
in which $\xi(t)$ is a Gaussian white noise verifying
\begin{equation}
 \label{eq:38} \langle \xi(t)\rangle=0\,, \quad \langle \xi(t) \xi(t')\rangle = 2 \alpha^{-1} \left[ \tilde{\mu}^2 \lambda'(x) \right]^2 \delta(t- t')\,.
\end{equation}
In Eq.~\eqref{eq:37}, the noise acts on the variable $p$
while $D(x)$ depends only on $x$, thus it is not multiplicative.

\textit{Numerical simulations.-} In order to check the consistency of
our theoretical scheme, we perform numerical simulations of the
``exact'' \textit{microscopic} dynamics \eqref{eq:4} in the
$\alpha \gg 1$--fast spins--limit: our aim is to compare the measured
values of significative observables to those predicted by the
\textit{mesoscopic} description provided by the Fokker-Planck
equation~\eqref{eq:32}. Specifically, we consider the following case
\begin{align}\label{eq:choices}
 K(p)=1-\cos p, \quad U(x)=\left(1-\cos x \right)^2, \quad
  \lambda(x)=\sin x\,.
\end{align}
The kinetic energy is inspired by the experiment in
Ref.~\cite{braun_negative_2013}, where cold atoms in an optical
lattice display both positive and negative temperatures. It has also
been studied theoretically, for instance
see~\cite{cerino_consistent_2015,baldovin_langevin_2018}.

For the microscopic dynamics, the spins are started from a completely
random configuration.  Then, for each time-step $dt$ thereof,
our algorithm performs two actions: first, it evolves the state
$(x,p)$ of the particle through a deterministic Velocity Verlet
integration step; then it chooses one spin with uniform probability,
and tries to flip it according to the Glauber
dynamics~\eqref{eq:3}. The probability of flipping the chosen spin
$\sigma_j$ is given by $N dt \,W_j(\bm{\sigma}|x)$; in order to
keep it of the order of unity, we choose $dt=(\alpha N)^{-1}$ for our
simulations.

As a first check of the validity of our description, we verify the
renormalisation of the potential that arises in our theoretical
framework. Specifically, we check the shape of the equilibrium PDF for
the particle variables $(x,y)$, which is given by
Eq.~\eqref{eq:marginal-prob-eq}. Making use of Eq.~\eqref{eq:21b} and
\eqref{eq:choices}, the renormalised potential $U_R(x)$ is
\begin{equation}
  U_{R}(x)=(1-\cos x)^{2}-\frac{\beta}{2}\tilde{\mu}^{2}\sin^{2}x\,.
\end{equation}
For positive temperatures, $U_{R}(x)$ corresponds to a bistable
potential with symmetric minima at $x\in[-\pi,\pi]$ verifying
$\cos x=2/(2+\beta\tilde{\mu}^{2})$ and maxima at $x=0,\pm\pi$,
whereas for negative temperatures $U_{R}(x)$ has only one minimum at
$x=0$ and attains its maximum value at $x=\pm\pi$. Thus, the most
probable value of $x$--given by the maximum of
$\exp[-\beta U_{R}(x)]$--changes discontinuously from
$x=\pm\tilde{\mu}\sqrt{\beta}$ for $\beta=0^{+}$ to $x=\pm\pi$ for
$\beta=0^{-}$.

\begin{figure}
\includegraphics[width=\linewidth]{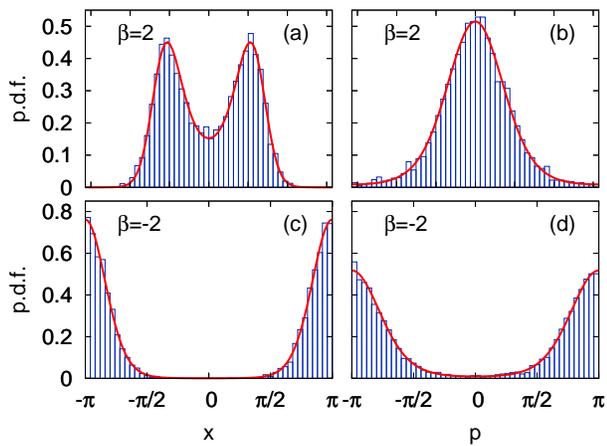}
\caption{Equilibrium PDF of the particle variables $(x,p)$. The top
  (bottom) panels correspond to $\beta=2$ ($\beta=-2$). Histograms are
  computed from numerical simulations of the microscopic dynamics
  \eqref{eq:4}; red solid lines are the best fits to the Boltzmann
  distribution~\eqref{eq:marginal-prob-eq}. Parameters: $N=10^4$,
  $\alpha=10$, $\mu=10^{-2}$, $dt=(\alpha N)^{-1}$.}
\label{fig:distr}
\end{figure}

In Fig.~\ref{fig:distr}, we show the histograms of $x$ and $p$ at
equilibrium, for two values of the temperature with opposite signs:
the agreement between the numerical and the theoretical results are
excellent.  By fitting each plot with the corresponding Boltzmann
factor, we infer values of the parameter $\beta$ that are compatible
with the original ones used in the simulations, within the confidence
interval for the fit. Note that the most probable value of
momentum is $\pm\pi \neq 0$ for $\beta<0$, but this is compatible with
stationarity: there is no average drift
since $\dot x = \partial H/\partial p =\sin p$.

\begin{figure}
\includegraphics[width=\linewidth]{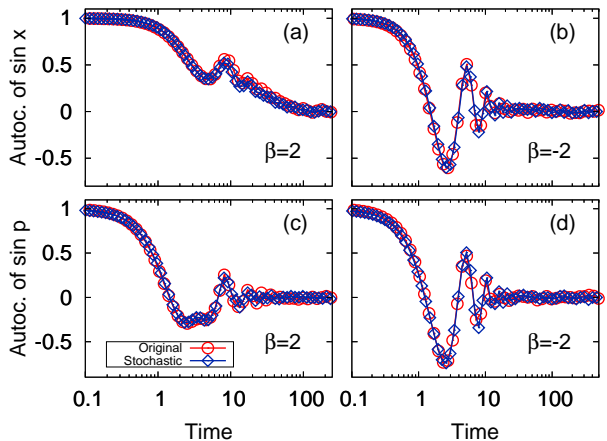}
\caption{Equilibrium autocorrelation functions. Specifically, we
  consider two observables, $\sin x$ above and $\sin p$ below, for
  $\beta=2$ (left panels) and $\beta=-2$ (right panels). Red circles
  represent the simulations of the original dynamics~\eqref{eq:4},
  whereas blue diamonds are the numerical integration of the
  Fokker-Plank equation~\eqref{eq:32}, with a time-step
  $h=10^{-4}$. Other parameters as in Fig.~\ref{fig:distr}.}
\label{fig:autoc}
\end{figure}

\begin{figure}
\includegraphics[width=\linewidth]{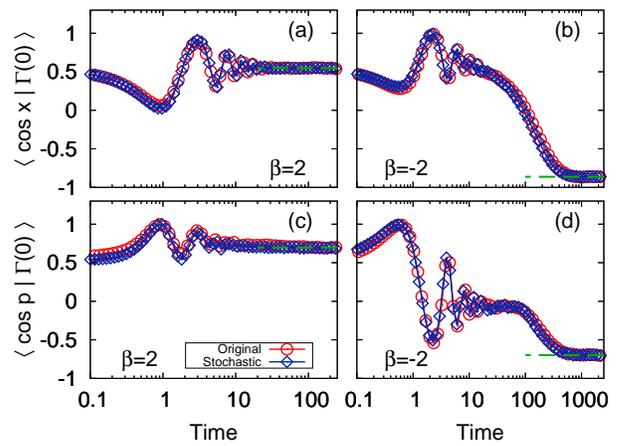}
\caption{Relaxation to equilibrium. We plot the time evolution of the
  averages of $\cos x$ (top) and $\cos p$ (bottom), for $\beta=2$
  (left panels) and $\beta=-2$ (right panels). In all cases, the
  particle starts from the initial condition
  $\Gamma(0)=(x(0),p(0))=(1,1)$. As in Fig.~\ref{fig:autoc}, the
  original dynamics (red circles) is compared with the Fokker-Planck
  equation (blue diamonds), with a time-step $h=10^{-4}$. Other
  parameters as in Fig.~\ref{fig:distr}. Green dashed lines are
  $\langle \cos x \rangle$ and $\langle \cos p \rangle$ analytically
  computed averaging over the theoretical equilibrium distributions.}
\label{fig:average}
\end{figure}

Second, we check the accuracy of the derived Fokker-Planck equation
for describing the dynamics of the particle variables. More
concretely, it is how the dynamical quantities obtained from
Fokker-Planck compare with those obtained from the exact dynamics that
we are interested in. With this aim, we numerically integrate
Eq.~\eqref{eq:37} using a standard algorithm for stochastic
differential equations \cite{mannella_fast_1989}--in its variant of
order $h^{3/2}$, where $h$ is the time-step.

Several time-dependent quantities computed from the Fokker-Planck
equation~\eqref{eq:32} are compared with those obtained by simulating
the original Liouville-master equation~\eqref{eq:4}.  In
Fig.~\ref{fig:autoc}, we look into time correlation functions at
equilibrium, namely, into the autocorrelations of $\sin x$ and
$\sin p$. The qualitative difference between panels (a) and (b) can be
related to the shape of the free energy, which is different for
positive and negative temperatures. When it is bi-stable ($\beta>0$),
the time needed to cross zero is longer and thus oscillations are
hindered. In Fig.~\ref{fig:average}, we study the relaxation to
equilibrium of some dynamical observables. In particular, we have
evaluated $\langle \cos x \rangle$ and $\langle \cos p \rangle$,
conditioned to fixed initial values of the particle variables
$\Gamma(0)\equiv (x(0),p(0))$. In both cases, the agreement is
evident.

\textit{Concluding remarks.-} In conclusion, we have generalised the
problem of deriving a LE to non-standard forms of
the Hamiltonian that also allow for absolute negative
temperatures. The LE obtained here satisfies a
generalised Einstein relation that has been shown to apply for (i)
arbitrary spatial dependence of the transport coefficients, and (ii)
situations in which the potential felt by the particle is renormalised
as a consequence of its interaction with the bath. Such a
renormalisation is relevant when the eliminated fast degrees of
freedom change the potential felt by the
particle~\footnote{For an application of these ideas to biomolecules
  and buckling in graphene
  see~\cite{prados_spin-oscillator_2012,ruiz-garcia_ripples_2015,
    ruiz-garcia_bifurcation_2017}.}.

A particular example is treated in detail through a
Chapman-Enskog-like coarse-graining procedure, which provides exact
expressions for the transport coefficients. This specific case is in
complete agreement with the general picture and, in addition, presents
a transition from one-basin to bi-stable free energy when going from
positive to negative temperatures.

\begin{acknowledgments}
  Antonio Prados would like to acknowledge financial support from the Spanish
  Ministerio de Ciencia, Innovaci\'on y Universidades through Grant
  No.~PGC2018-093998-B-I00 (partially financed by the ERDF).
\end{acknowledgments}


\end{document}